\documentclass[aip,pop,sd,reprint,10pt,amsmath,amssymb,graphicx]{revtex4-1}

\usepackage{bm}
\usepackage{dcolumn}
\usepackage{graphicx}
\usepackage{color}
\usepackage{caption}
\usepackage{subcaption}
\usepackage{soul}
\usepackage{siunitx}
\usepackage{verbatim}
\usepackage{amsmath}
\usepackage{mathtools}
\usepackage{amssymb}

\newcommand{\nc}{\newcommand}
\nc{\rcite}[1]{Ref.~\onlinecite{#1}}
\nc{\rcites}[1]{Refs.~\onlinecite{#1}}
\nc{\eqeqref}[1]{Eq.~\eqref{eq:#1}}
\nc{\eqseqref}[2]{Eqs.~\eqref{eq:#1}-\eqref{eq:#2} }
\nc{\secref}[1]{Sec.~\ref{sec:#1}}
\nc{\secsref}[2]{Sec.~\ref{sec:#1}-Sec.~\ref{sec:#2}}
\nc{\ssecref}[1]{Sec.~\ref{ssec:#1}}
\nc{\ssecsref}[2]{Sec.~\ref{ssec:#1}-Sec.~\ref{ssec:#2}}

\begin{document}
\title{Analysis of equilibrium and turbulent fluxes across the separatrix in a gyrokinetic simulation}
\author{I. Keramidas Charidakos}
\affiliation{University of Colorado Boulder, Boulder}
\author{J. R. Myra}
\affiliation{Lodestar Research Corporation, 2400 Central Avenue, Boulder, Colorado 80301, USA}
\author{S. Parker}
\affiliation{University of Colorado Boulder, Boulder}
\author{S. Ku}
\affiliation{Princeton Plasma Physics Laboratory}
\author{R.M. Churchill}
\affiliation{Princeton Plasma Physics Laboratory}
\author{R. Hager}
\affiliation{Princeton Plasma Physics Laboratory}
\author{C.S. Chang}
\affiliation{Princeton Plasma Physics Laboratory}

\begin{abstract}
The SOL width is a parameter of paramount importance in modern tokamaks as it controls the power density deposited at the divertor plates, critical for plasma-facing material survivability. An understanding of the parameters controlling it has consequently long been sought (Connor \textit{et al}. 1999 NF \textbf{39} 2). Prior to Chang \textit{et al.} (2017 NF \textbf{57} 11), studies of the tokamak edge have been mostly confined to reduced fluid models and simplified geometries, leaving out important pieces of physics. Here, we analyze the results of a DIII-D simulation performed with the full-f gyrokinetic code XGC1 which includes both turbulence and neoclassical effects in realistic divertor geometry. More specifically, we calculate the particle and heat $E\times B$ fluxes along the separatrix, discriminating between equilibrium and turbulent contributions. We find that the density SOL width is impacted almost exclusively by the turbulent electron flux. In this simulation, the level of edge turbulence is regulated by a mechanism we are only beginning to understand: $\nabla B$-drifts and ion X-point losses at the top and bottom of the machine, along with ion banana orbits at the low field side (LFS), result in a complex poloidal potential structure at the separatrix which is the cause of the $E\times B$ drift pattern that we observe.  Turbulence is being suppressed by the shear flows that this potential generates. At the same time, turbulence, along with increased edge collisionality and electron inertia, can influence the shape of the potential structure by making the electrons non-adiabatic. Moreover, being the only means through which the electrons can lose confinement, it needs to be in a balance with the original direct ion orbit losses to maintain charge neutrality.

\nocite{connor1999comparison, goldston2011heuristic}
\end{abstract}
\maketitle

\section{INTRODUCTION}
Studies of the plasma edge are indispensable for a variety of tokamak operation and performance issues such as the density and power scrape-off-layer (SOL) widths\cite{chang2017gyrokinetic,meier2017drifts,meier2016analysis,myra2011reduced,russell2015modeling,pankin2015kinetic,rozhansky2018structure,reiser2017drift}, intrinsic rotation and momentum transport\cite{stoltzfus2012tokamak,stoltzfus2012transport,muller2011experimental,loizu2014intrinsic,groebner2009intrinsic,chang2008spontaneous,seo2014intrinsic}, edge flows\cite{chankin2007possible,hoshino2007numerical,kirnev2005edge2d,labombard2004transport,pigarov2008simulation,pitts2005edge}and the L-H transition\cite{aydemir2012pfirsch,shaing1989bifurcation,chang2017fast}. 

Out of all the above mentioned important and inter-related facets of edge physics, the SOL width is crucial in tokamak operation as it governs the power density delivered at the divertor plates. Indeed, a viable fusion reactor would need to have plasma-facing materials that can withstand large heat loads without the need of regular replacement. However, an ITER exhaust power of $\SI{100}{\mega\watt}$, concentrated on a speculated $\sim \SI{1}{\milli\meter}$ SOL width would produce a whopping $\SI[per-mode=fraction]{5.5}{\giga\watt\per\square\meter}$ of parallel power density, placing stringent constraints on the lifetime of the divertor materials\cite{goldston2015theoretical}. It is therefore of utmost importance to be able to predict the SOL width for a given discharge based on its parameters but also to gain an understanding of the physical mechanisms that control its size. To this end, Eich\cite{eich2013scaling} has constructed scaling relations for the power SOL width, $\lambda_q$, based on multiple regressions of relevant parameters from various experimental discharges. He has arrived at a scaling law of the form $\lambda_q\sim B^{-1.19}_{p}$, where $B_p$ is the poloidal magnetic field, based on which the SOL width of ITER is inferred to be $\SI{1}{\milli\meter}$. Goldston\cite{goldston2011heuristic} has provided theoretical justification for the above scaling by devising a heuristic model that predicts the SOL width. In summary, this model rests on the assumption that the perpendicular $\nabla B$ and curvature drift ion flows are balanced exactly by parallel flows with one half of the flux being directed to the divertors at a speed of $\frac{C_s}{2}$ and the other half returning upstream via the Pfirsch-Schl{\"u}tter current. This mechanism is responsible for setting the density SOL width, $\lambda_n$. The next assumption is that this density channel fills with turbulent electrons and empties through Spitzer-H{\"a}rm conductivity, establishing the power SOL width. The heuristic model correctly captures the inverse relation between $\lambda_q$ and $B_p$ proposed by Eich and can thus, with reasonable accuracy, reproduce the experimental results.  

Until recently, simulations of the edge have relied upon reduced fluid models in simplified geometries omitting essential physics. This changed with the introduction of XGC1\cite{chang2009compressed,ku2009full,ku2016new,churchill2017pedestal,chang2017fast} which is a full-f, Particle-In-Cell, 5D gyrokinetic code using realistic X-point geometry, optimized to simulate the tokamak edge. The code has gyrokinetic ions and offers an option for drift-kinetic, fluid or adiabatic electrons. Employing it in a study of ITER SOL width, Chang et.al\cite{chang2017gyrokinetic} discovered that the simulated width is six times larger than the one predicted by the Eich fit or the Goldston heuristic model, casting doubt on whether the heretofore assumed scaling relations continue to be valid at the ITER regime. The explanation that they provide regarding this disparity  has to do with the importance of edge electron turbulence\cite{d2011convective,myra2015turbulent,myra2016theory} in the new regime. More specifically, they claim that the electrostatic-`blobby' edge turbulence is setting the scale of the electron channel. The size of a blob or a streamer (or any other coherent turbulent structure) is, at least partially, related to the width of the edge shearing layer\cite{furno2008experimental,bisai2005formation}, wherein the blobs are being generated. In present day tokamaks, this shearing layer width is relatively small and the density channel is dominated by ion neoclassical flows inside the SOL. On ITER however, due to the much larger size of the shearing layer, the radial extent of blobs and streamers is bigger and turbulence dominates the density width.  

From the above discussion, it is clear that important figures of merit such as the SOL width are the result of a non-linear interaction between particle orbit effects (e.g., neoclassical effects and ion X-point losses\cite{chang2002x,ku2004property}) with the electrostatic-`blobby' edge turbulence. The interaction is through the radial and poloidal edge electric field, established by the ion losses and the sheared radial flows that it generates which can influence both the turbulence and the ion flows. Such interplay can often yield unexpected results and thus merits further study with state-of-the-art codes which might illuminate certain aspects of this complex process. Here, we analyze the results of an XGC1 simulation of DIII-D from the viewpoint of equilibrium and turbulent $E\times B$ particle and heat fluxes across the last closed flux surface (LCFS). Within this framework, we attempt to understand the relationship between the direct ion orbit losses and the electrostatic potential that drives these fluxes as well as the effect of the generated shear on the turbulence. Furthermore, we try to measure the impact of turbulence on the electron transport and thereby, test the second assumption of the heuristic model which states that only turbulent electrons are responsible for filling the density SOL channel.

The organization of the paper is as follows: In Section \ref{sec:simulation} we give a brief description of the simulation, it's geometry and some relevant parameters. In Section \ref{sec:fluxes} we provide the definitions of the measured fluxes and the methodology of extracting them from the XGC1 data. We present the equilibrium and turbulent $E\times B$ particle and heat fluxes across the separatrix and interpret the observed patterns using the notions of ion X-losses, ion banana drifts, non-adiabaticity, quasineutrality and shear flows. In Section \ref{sec:numerical} we perform numerical integrations to assess the size of the different fluxes and extricate the contribution of each to the cross-separatrix transport and hence, to the SOL width. In Section \ref{sec:conclusions} we summarize and draw conclusions.  

\section{Simulation}\label{sec:simulation}
\begin{figure}[h]
\centering
\includegraphics[scale=1.3]{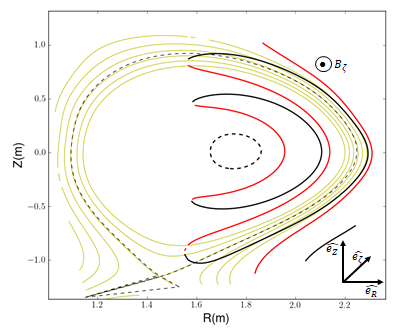}
\caption{Poloidal cross-section of the simulation domain with separatrix, flux surfaces and banana orbits highlighted. }\label{geometry}
\end{figure}

In this section we are going to provide a description of the simulation domain and parameters. The simulation is of the DIII-D\cite{luxon2002design} discharge \# 144981, which is an H-mode heated by neutral beam injection. It is initialized with experimental profiles taken at time $\SI{3175}  {\milli\second}$. The simulation inputs include experimental kinetic profiles of electron density and temperature ($n_e$ and $T_e$), ion temperature ($T_i$), and magnetic equilibrium, from a kinetic EFIT magnetic reconstruction. In Fig.~\ref{geometry} we see a poloidal cross-section of the simulation domain along with the directions of the unit vectors of the cylindrical coordinate system $(R,\zeta,Z)$ used in this paper. The origin of the coordinate system in the poloidal cross-section will be taken to be the point $(R,Z) = (R_o, Z_o) = (\SI{1.67}{\meter},0)$. The geometry is lower single null but with a secondary ``virtual'' upper X-point (outside of the simulation domain). The magnetic field is in the negative $\hat{\zeta}$ direction making the ions $\nabla B$-drift towards the X-point and the electrons towards the top. The black dashed line denotes the separatrix. The yellow lines are closed and open flux surfaces and the black and red continuous lines are showing the poloidal projection of the banana orbits (black for ascending-red for descending parts of the orbit). Those have been plotted using contours of toroidal angular momentum. The closed dashed black line close to the center is a potato orbit. The orientation of the coordinate system and the toroidal magnetic field direction are also shown. In the rest of the paper, when we refer to the poloidal angle $\theta$, of a separatrix point, we will mean the angle formed between the line passing through that point and the line passing through the low-field side midplane with both lines passing through the origin. Positive angles are in the counter-clockwise direction. 

The total simulation time is $\SI{0.16}{\milli\second}$ with a time-step of $\num{2.3d-4}\si{\milli\second}$. Although this time is not enough for the core turbulence to saturate it is adequate for the saturation of the edge turbulence. The poloidal magnetic field is $B_{pol} = \SI{0.42}{\tesla}$ (measured at the outboard midplane, in the positive $\hat{e}_{\theta}$ direction) and the toroidal current is $I_{p} = \SI{1.5}{\mega\ampere}$. Equilibrium plasma density at the edge was measured to be $n\approx \SI{3.5e19} {\meter^{-3}}$ with a Greenwald density of $n_G\approx \SI{15.6e19}{\meter^{-3}}$. Sound speed and poloidal flow were estimated from the data to be $C_s\approx \SI[per-mode=fraction]{15e4}{\meter\per\second}$ and $V_{\theta} \approx \SI[per-mode=fraction]{4e4}{\meter\per\second}$ respectively. Safety factor close to the edge was found to be $q_{95} =  3.7$ giving a connection length of about $qR\approx \SI{8.6}{\meter}$. This connection length, along with an electron thermal velocity of $u_{te}\approx \SI[per-mode=fraction]{5.1e6}{\meter\per\second}$ results in a transit frequency of $\SI{593}{\kilo\hertz}$. The collision times were calculated at $\tau_{ei} = \SI{3.5}{\micro\second}$ and $\tau_{ii} = \SI{0.5}{\milli\second}$. The ions and electrons are weakly collisional thus, the adiabatic invariant $\mu$ is conserved over many bounce times. 

Given that in the next section, the trapped particle orbits are going to play a fundamental role in our interpretation of the flux patterns, here, we give a few relevant numerical estimates regarding those. The fraction of trapped particles at the edge was estimated from the relation $\sqrt{\epsilon} (1.46-0.46 \epsilon) \approx 0.767$ with $\epsilon$ being the aspect ratio. The bounce times of deeply trapped particles were calculated to be $\tau_{b,i} \approx \SI{0.16}{\milli\second}$ and $\tau_{b,e} = \SI{6}{\micro\second}$. Technically, the bounce time of a deeply trapped particle whose banana orbit turning point is located at the high field side (HFS) midplane is infinite. The above bounce times are simple estimates found by dividing the length of a poloidal circuit with the magnetic drift speed of each particle and they are presented here to make the point that almost all of the ions have enough time, within the simulation, to complete a banana orbit. With $\epsilon^{3/2}\approx 0.2$, we calculate $\nu_{\star i} = \frac{\nu_{ii} q_{95} R}{u_{ti}} \approx 0.12$ and $\nu_{\star e} = \frac{\nu_{ei} q_{95} R}{u_{te}} \approx 0.48$, we can deduce that at the edge, the electrons are in the plateau regime ($\epsilon^{3/2} < \nu_{\star e} < 1$) whereas the ions are in the banana transport regime ($\nu_{\star i} < \epsilon^{3/2}$).



\section{The Fluxes}\label{sec:fluxes}
\subsection{Flux definitions}

One of the goals of the present paper is to explore the relationship between equilibrium and turbulent processes and the associated fluxes across the separatrix.  For dynamical quantities such as the plasma density $n$, or velocity $v$ we define:

\begin{align}
    n &= \left<n\right>_{t,\zeta} + \delta n \,,\notag\\
    v &= \left<v\right>_{t,\zeta} + \delta v \,,\label{turb_def}
\end{align}

where $\left<\cdots\right>$ denotes an average in either time $(t)$ or toroidal planes $(\zeta)$, or both. The time averages considered here are taken over a time interval late in the simulation where a quasi-steady turbulent state has been achieved.

Employing Eq.~\eqref{turb_def}, the fundamental relationship for the fluxes is:
\begin{equation}
    \left<n v\right>_{t,\zeta} = \left<n\right>_{t,\zeta}\left<v\right>_{t,\zeta} + \left<\delta n \delta v\right>_{t,\zeta}\,,\label{flux_def}
\end{equation}
where the cross terms $\left<n \delta v\right>_{t,\zeta}$ and $\left<v \delta n\right>_{t,\zeta}$ vanish due to the vanishing of $\left<\delta n\right>_{t,\zeta}$ and $\left<\delta v\right>_{t,\zeta}$.

These ``fluxes" are not technically transport fluxes but rather local density weighted flows. Although collisions are fully accounted for in the simulation, the rapid electron transit time implies that the radial excursions of electrons due to either magnetic or $E\times B$ drifts are small and can cancel without causing net transport as is well known from neoclassical theory. In this paper, we will continue to refer to them as fluxes for brevity. As will be seen, they provide a useful diagnostic for understanding basic properties of both the background plasma and the resulting turbulence. Often in the literature, we find the first piece of the rhs of Eq.~\eqref{flux_def} to be called the equilibrium flux whereas the second piece is known as the turbulent flux. In this work, we employ the same definition for the turbulent flux however, for convenience in analysis the definition of the equilibrium flux employed in this paper is slightly modified. Our equilibrium flux is given by $\Gamma_{eq} = \left<\left<n\right>_{\zeta}\left<v\right>_{\zeta}\right>_{t}$. These two definitions of the equilibrium flux are not exactly equivalent. They are connected by the relation: $\left<\left<n\right>_{\zeta}\left<v\right>_{\zeta}\right>_{t} = \left<n\right>_{t,\zeta}\left<v\right>_{t,\zeta} - \left<\left<\delta n\right>_{\zeta}\left<\delta v\right>_{\zeta}\right>_{t}$\,. However, independent calculation of the difference $\left<\left<\delta n\right>_{\zeta}\left<\delta v\right>_{\zeta}\right>_{t}$\,, has shown that this ``ringing" term, associated with temporal fluctuations of the toroidally averaged fields, is three orders of magnitude smaller than the two equilibrium fluxes indicating that, for the level of numerical accuracy assumed in this work, we can take the two definitions to be equivalent. 

The definitions for the heat fluxes are similar, with the replacement of the density by the pressure of each species. Because the simulation is quasineutral, we can not distinguish between the densities or the net particle fluxes of the two species. In general both magnetic and ExB drifts contribute to the local particle fluxes crossing the separatrix at a particular poloidal position.  In principle, the electron and ion ExB drift fluxes can be different due to the larger ion gyroradius. This orbit averaging effect results in the suppression of the ion $E \times B$ fluxes by a factor $\Gamma_o(b = k^2_{\perp}\rho^2_i) = e^{-b}I_o(b)$. Close to the separatrix, $\rho_i \approx \SI{2}{\milli\meter}$ and from power spectra of equilibrium and turbulent fields, $k^{eq}_{\perp}\approx \SI{61}{\meter^{-1}}$, $k^{tur}_{\perp}\approx \SI{123}{\meter^{-1}}$ which gives a suppression of about $1\%-2\%$ for the equilibrium and about $5\%-6\%$ for the turbulent $E \times B$ fluxes. In the case of the heat fluxes though, a clear distinction between ions and electrons can be made, based on their different temperatures. Finally, we should note that, given the available diagnostics, we lack the capability of calculating the turbulent polarization current and its flux. Nevertheless, in later sections, we estimate its magnitude relative to the rest of the fluxes.   

\subsection{Flux calculation}
In this section, we describe our method for calculating the various fluxes, using the XGC1 data, which illustrates the limitations of the particular data set and can be useful to practitioners in the field. 

The calculation of radial $E \times B$ drift velocities from the density and potential fields, whose values we know on the computational nodes, involves derivatives along the $R$ and $Z$ directions. However, a plot of time and toroidally averaged $n_e$ and $\Phi$ on the separatrix as the one in Fig.~\ref{adiabaticity} shows that, despite the fact that a characteristic trend can be immediately identified, the values are noisy, especially at the low field side (LFS). Therefore, the strategy of interpolating the computational node values along horizontal and vertical lines and then taking derivatives along those lines to construct the vector velocity can give highly inaccurate results particularly for a sub-dominant component, such as the radial component of the $E \times B$ velocity. For that reason, we resorted to expressing the required derivatives as derivatives with respect to the poloidal angle $\theta$. This can be accomplished by writing the magnetic field in terms of the coordinate system $(\Psi,\zeta,\theta)$. In this coordinate system, the magnetic field is $B = R B_{\zeta} \nabla \zeta + \nabla \Psi\times \nabla \zeta$. Using the formula $\mathcal{J}^{-1} = \nabla \Psi \cdot \nabla \theta \times \nabla \zeta$ for the inverse Jacobian, we can turn the expression for the radial component of the $E\times B$ velocity, $V_{E_\Psi} = \hat{e}_{\Psi}\cdot \frac{\hat{b}\times \nabla \Phi}{B}$, into $$V_{E_\Psi} = -\left(\frac{1}{B^2 R B_p }\right)\left(\frac{R B_{\zeta}}{\mathcal{J}}\frac{\partial \Phi}{\partial \theta} - B^2_p \frac{\partial \Phi}{\partial \zeta}\right)\,.$$ For all equilibrium quantities, $\frac{\partial \Phi_{eq}}{\partial \zeta} = 0\,,$ since the system has axisymmetry. For turbulent quantities, we can compute such derivatives using the assumption that turbulent structures are field aligned. Demanding $B\cdot \nabla \delta\Phi\approx 0\,,$ we arrive at the equation $$\frac{\partial \delta \Phi}{\partial \zeta} = \frac{R}{B_{\zeta}\mathcal{J}}\frac{\partial \delta \Phi}{\partial \theta}\,,$$ which can be readily used. The above expressions allow us to work with derivatives in the poloidal direction for the calculation of the $E\times B$ velocity. This procedure has the advantage that we only need to do a single interpolation and smoothing of $n_e$ and $\Phi$ around the LCFS and then take the derivative. The results thus obtained are much less noisy.

\subsection{The Particle Fluxes}

First, we will describe the fluxes qualitatively and attempt to give an intuitive explanation for their shapes. Afterwards, we will seek to arrive at relevant quantitative conclusions based on their relative sizes.

\subsubsection{Equilibrium Flux}
We start with the spatial distribution of the equilibrium, $E\times B$ particle flux normal to, and hence crossing, the separatrix, which is common between the two species, as discussed above.

\begin{figure}[h]
\centering
\includegraphics[scale=0.5]{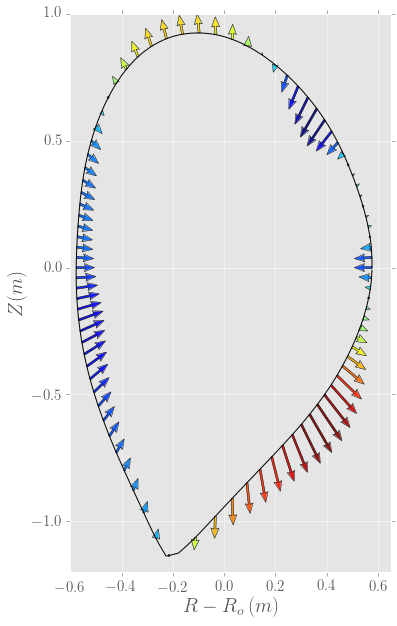}
\caption{Equilibrium radial $E\times B$ particle flux.}\label{Particle_Eq}
\end{figure}

In Fig.~\ref{Particle_Eq} we observe that the magnitude of the particle flux (arrow length and color indicate flux intensity) is not uniform and exhibits a strong poloidal variation, not all of which can be explained by the difference in field strength between the inboard and outboard sides. Certainly, the flux is suppressed at the HFS compared to the LFS. However, at the LFS, the flux seems to be concentrated at two lobes, one outward, starting near the X-point and extending up to a little below the midplane, and one inward, which starts above the midplane and extends near the top of the machine. At the top, near the top X-point, the flux becomes again slightly outward, before it turns again to slightly inward at the HFS all the way to the bottom X-point.  

\begin{figure}[h]
\centering
\includegraphics[scale=0.6]{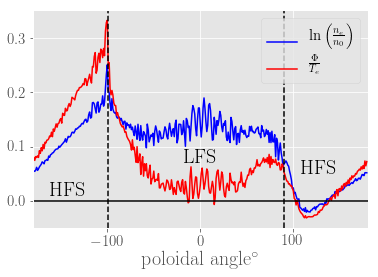}
\caption{Degree of adiabaticity between $\Phi$ and $n$.}\label{adiabaticity}
\end{figure}

To understand the poloidal pattern we just described, we turn to the distributions of density and potential on the LCFS which we show in Fig.~\ref{adiabaticity}. There, we plot the two terms of the adiabatic relation so that we get a measure of the electron departure from adiabaticity at various locations along the LCFS. Moreover, the $\frac{\phi}{T_e}$ term has the same structure as the potential itself, given that the electrons are almost isothermal on the flux surfaces. From this plot, we immediately recognize the points where the global minimum ($\phi \approx 120^{\circ} $) and global maximum ($\phi \approx 270^{\circ} $) are located, which are the top and bottom X-points respectively. Although the exact location of these two extreme points is a result of a complex interplay between Pfirsch-Schl{\"u}tter flows and collisional effects \cite{aydemir2012pfirsch,aydemir2009intrinsic}, it follows from the direction of the magnetic drifts: ions drift downwards and electrons upwards therefore, the potential arranges itself so that it attracts ions from the core at the top and electrons at the bottom. In both cases, the extremization of the potential, follows an analogous increase/decrease of density. In the case of the bottom X-point, an important factor that would lead to a density build-up, is the X-point ion loss. The X-point loss\cite{chang2002x,ku2004property} of hot ions leads to a lower ion temperature at the LCFS. It should be noted that, in our simulation, the $E_r$ well from the X-loss has already been established. Therefore, the loss cone has moved to higher energies, resulting in hot ion loss. Although the scalar ion pressure is not constant along a field line\cite{churchill2017total}, approximate pressure balance is consistent with the lower temperature to be offset by a density increase. These two extremal points of potential establish an electric field in the HFS region. It is this electric field which in turn creates the $E\times B$ flow which is responsible for the inward flux pinch we see in Fig.~\ref{Particle_Eq} on the HFS as well as the dominant outward flux on the LFS. The reason why this flux is not as high as the ones we see on the LFS, even if the electric field is stronger, is the mitigating effect of the strong magnetic field.      

In addition to the global minimum and the global maximum of $\phi$, we also have two local ones. The positions of those, correlate with the two bulges of opposite direction that we see on the LFS in Fig.~\ref{Particle_Eq}. Indeed, the directions of the fluxes are completely in line with the electric field directions established by the alternating pairs of minimums and maximums of $\phi$. However, in contrast to the two global extremal points of potential, it is not immediately clear what is responsible for the two local ones in the LFS. Here, we propose that this potential structure is produced by the orbits of trapped ions which, as we saw in the previous section, constitute roughly $76\%$ of the total ion population near the edge. A trapped ion moving on its banana orbit exits the separatrix at some point below the LFS midplane, leaving behind it a negative charge. If the electrons were completely adiabatic, negative charge would be instantly neutralized by electrons moving along the field lines to prevent the potential build-up. But, as we can readily verify from Fig.~\ref{adiabaticity}, there is a significant departure from adiabaticity at the LFS. Therefore, a trapped ion leaving the separatrix will create a negative potential in order to attract more ions. Conversely, the region above the LFS midplane is the region where the trapped ions re-enter the LCFS (at least those that didn't get entrained in the parallel SOL flow) generating a positive charge. The non-adiabaticity of the electrons, forces the potential to increase, in order to attract them at that location. Thus neoclassical ion orbits combined with X-point losses are the dominant mechanism responsible for the structure of the poloidal electric field, and hence the equilibrium radial particle flux in this simulation.

The non-adiabaticity of the electrons at the separatrix, seems to be crucial for the exact shape of the edge potential. This non-adiabaticity at the edge can be understood if we compare the time rates of all terms in the electron momentum equation. More specifically, the inertial term, i.e., the transit frequency, was given in Section \ref{sec:simulation} as $\SI{593}{\kilo\hertz}$. Although large, it is comparable to the electron collision frequency (also given in \ref{sec:simulation}) and the turbulent frequency, which is discussed in the next section and it is found to be $\SI{640}{\kilo\hertz}$. Therefore, both collisions and the inertial term are important in the electron momentum equation leading to the observed departure from adiabaticity.

\subsubsection{Banana Tip distribution}
The subject of neoclassical direct ion orbit losses and their contribution to the edge flows and electric fields has been treated substantially, both computationally and theoretically\cite{degrassie2015thermal,battaglia2014kinetic,miyamoto1996direct,stacey2015distribution,stacey2011effect,stacey2016recent,stacey2013interpretation,wilks2017calculation,wilks2016improvements,hahn2005wall}. Here, we offer a new argument to support our claim that the potential structure at the LFS is due to the excursions of trapped ions on their banana orbits. Assuming that the trapped particles, with orbits within a banana width from the edge, will create either a charge excess or a charge hole at the point of inflection (``banana tip''), we give an estimate for the angular distribution of those banana tips at the LFS.   

To get the fraction of trapped particles with banana tips in the range $(0,\theta)$, where $\theta$ is the angle that the magnetic field is $B$ and zero is the angle at $B_{min}$ (LFS midplane), we perform the following integral:

\begin{align}
f_t(B) &= \frac{1}{\sqrt{2\pi} v^3_t}\Sigma_{\sigma}\int d\!E \,e^{-\frac{E}{v^2_t}} \int^{\mu = E/B_{min}}_{\mu=E/B} d\!\mu\, \frac{B_{min}}{\sqrt{2}\sqrt{E - \mu B_{min}}} \,,\notag\\
f_t(B) &= \frac{1}{\sqrt{2\pi} v^3_t}\Sigma_{\sigma}\int d\!E \,e^{-\frac{E}{v^2_t}}\sqrt{2E}\left(1-\frac{B_{min}}{B}\right)^{\frac{1}{2}}\,,\notag\\
f_t(B) &= \frac{1}{2}\Sigma_{\sigma} \left(1-\frac{B_{min}}{B}\right)^{\frac{1}{2}}\,,
\end{align}
where we have changed coordinates from velocity space to the $E$, $\mu$ space, with $E$ being the energy, $\mu$ the adiabatic invariant. Here, $\sigma$ denotes the sign of the particles parallel velocity.

Plugging in the angular formula for $B$, $B = \frac{B_o}{1+\epsilon \cos\theta}$, valid for circular flux surfaces, we turn this into the cumulative angular distribution for banana tips:
\begin{equation}
    X(\theta) = \frac{1}{2}\Sigma_{\sigma}\sqrt{\frac{\epsilon}{1+\epsilon}}\sqrt{1-\cos\theta}\,.
\end{equation}

Interpreting $X(\theta)$ as $X(\theta) = \int^{\theta}_0 d\theta\, g(\theta)\,,$ we get $g(\theta) = \frac{d X(\theta)}{d\theta}\,.$ Therefore,

\begin{equation}
    g(\theta) = \frac{1}{2}\sqrt{\frac{\epsilon}{1+\epsilon}}\frac{\sin\theta}{\sqrt{1-\cos\theta}}\,.
\end{equation}

\begin{figure}[h]
\centering
\includegraphics[width = \linewidth]{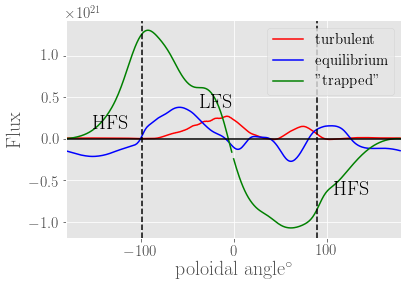}
\caption{Comparison of $E\times B$ fluxes with ``trapped" particle flux.}\label{trapped}
\end{figure}

The poloidal distribution of the flux caused by these particle excursions is the product of the tip distribution with the actual magnetic flux. In Fig.~\ref{trapped} we can see this flux compared to the $E\times B$ fluxes. We point out that indeed, the ``trapped" flux, tracks qualitatively the equilibrium $E\times B$ flux. Two things should be noted: the modelling of the tip distribution is rather simple as it assumes circular flux surfaces and completely ignores the effect of the X-point and the ion losses there. Also, a qualitative similarity between the fluxes is all we can hope for. Indeed, if the equilibrium $E\times B$ flux pattern at the LFS is the result of the potential structure due to the trapped particle excursions, as we claim, then, the relationship between their relative sizes is far from obvious.

\subsubsection{Turbulent Flux}
\begin{figure}
\includegraphics[width = \linewidth]{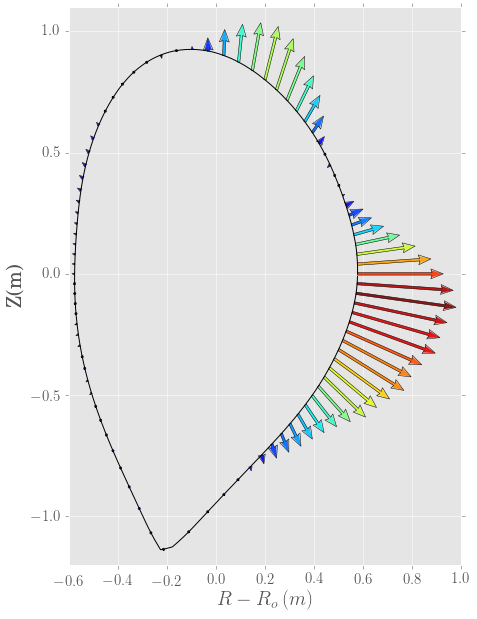}
\caption{Poloidal pattern of turbulent radial $E\times B$ flux.}
\label{tur_flux}
\end{figure}

We proceed with the description of the turbulent $E\times B$ flux which we show in Figs.~\ref{trapped} and ~\ref{tur_flux}. There, we observe that particle turbulence is entirely confined to the LFS and has a ballooning-like shape which peaks near the midplane, gets interrupted above it up to roughly $50^{\circ}$, picks up again after that and recedes off at the top of the machine.

\begin{figure}
\includegraphics[width = \linewidth]{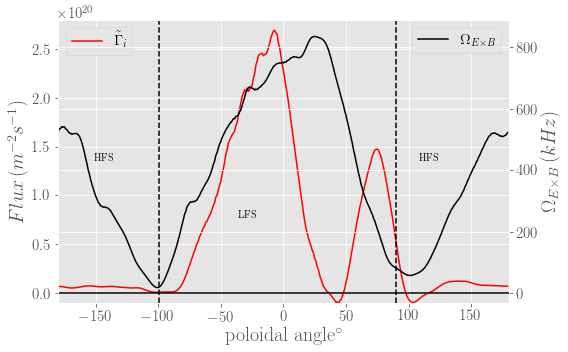}
\caption{Shear rate vs. turbulent flux.}
\label{shear_tur}
\end{figure}

A probable explanation for such a shape is provided by Fig.~\ref{shear_tur}. There, we overlay the plots of the turbulent flux and the $E\times B$ shearing rate on the LCFS. We observe that as the shearing rate exceeds a certain value (roughly at $640 kHz$) the turbulent flux drops sharply. When the $\Omega_{E\times B}$ drops below this value, the turbulence raises again. Indeed, we have measured the main frequency of the turbulence in this discharge to be very close to $\SI{640}{\kilo\hertz}$. We can imagine therefore, that in the absence of shear, the turbulence would have created a ballooning structure in the LFS, peaking at the midplane and tapering off at the top, where the curvature drive is minimal. However, the presence of shear at exactly the right level, suppresses it locally creating the resulting pattern\cite{burrell1997effects,hahm1995flow,terry2000suppression,biglari1990influence}. 

Recall from the previous discussion that the electrostatic potential pattern, and hence the shear, appears to be controlled by neoclassical drift-orbit effects. It is notable that the system has arranged itself so that this equilibrium shear is of sufficient magnitude to influence the turbulence.  We have verified that turbulence generated Reynolds stresses play a negligible role in establishing these sheared flows, which is not unexpected for this H-mode phase of the discharge\cite{muller2011experimental}.  

In addition to shear, it is possible that the observed turbulence pattern may also be related to the existence of curvature driven modes which peak at the top of torus\cite{abdoul2017generalised, halpern2013ideal}. Poloidal flux surface expansion, i.e., the distance between adjacent flux surfaces which is increasing towards the top of the machine, may also play a role.  First, it increases the outward velocity and hence the $ \Gamma $ of filamentary turbulent structures in order to maintain their field-aligned character\cite{galassi2017drive}. Second, flux surface expansion reduces $E\times B$ shear more quickly than the density or temperature gradients, since the former are proportional to a second radial derivative of the potential and therefore scale more strongly than the gradients which are proportional to a single radial derivative of the profiles.

Lastly, given the turbulent frequency, we can evaluate the relative size of the ion polarization flux compared to the turbulent $E\times B$ flux. Their ratio is approximately $\frac{\omega_{tur}}{\omega_{ci}} \approx 0.04$, which means that our inability to calculate the polarization current does not threaten the credibility of our results, at the present level of precision.

\subsection{The Heat Fluxes}
\begin{figure}
\includegraphics[width = \linewidth]{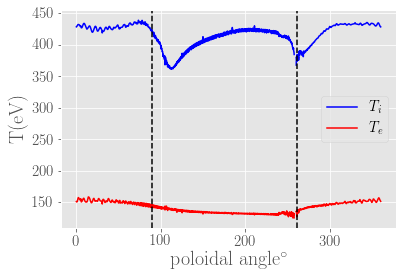}
\caption{Ion and Electron temperatures.}
\label{temperatures}
\end{figure}

In this section we present the qualitative features of the heat fluxes. Contrary to the $E\times B$ particle fluxes where we couldn't distinguish between the two species (apart from the, almost negligible, gyroaveraging effect of the ion orbits), for the heat fluxes we have a clear separation due to the different temperatures, as shown on Fig.~\ref{temperatures}. Therefore, in what follows, the calculation of the fluxes still assumes quasineutrality but each of the species has its own temperature. In calculating the heat fluxes, we employ Eq.~\eqref{flux_def} where in the place of velocity we substitute the pressure $P_s = n T_s$, with $s$ denoting the species. Therefore, the heat flux is given by the formula

\begin{equation}
    \left<P v\right>_{t,\zeta} = \left<P\right>_{t,\zeta}\left<v\right>_{t,\zeta} + \left<\delta P \delta v\right>_{t,\zeta}\,,\label{heat_def}
\end{equation}
where again, we call the first term of the rhs the ``equilibrium" and the second the ``turbulent" heat flux. The velocity here is the $E\times B$ one.

\begin{figure}
\includegraphics[width = \linewidth]{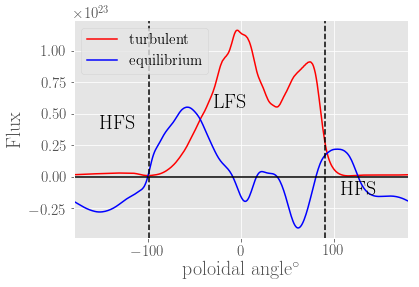}
\caption{Equilibrium and turbulent $E\times B$ heat fluxes of electrons.}
\label{el_heat}
\end{figure}

\begin{figure}
\includegraphics[width = \linewidth]{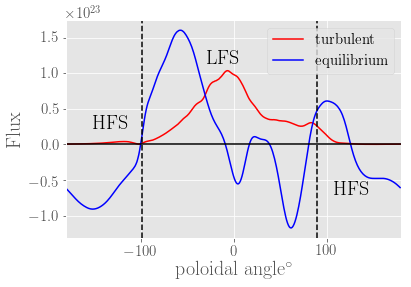}
\caption{Equilibrium and turbulent $E\times B$ heat fluxes of ions.}
\label{ion_heat}
\end{figure}

In Figs.~\ref{el_heat}-\ref{ion_heat} we present the equilibrium and turbulent $E\times B$ fluxes of electrons and ions respectively, along the LCFS. The shapes of both species equilibrium flux is similar except for the scaling effect of the temperature difference. This can be easily understood from Fig.~\ref{temperatures} where we plot the temperatures as a function of the poloidal angle. There, we see that the electrons are very close to being isothermal. Therefore, the qualitative features between the particle equilibrium and electron equilibrium $E\times B$ heat fluxes are common. 

The ion temperature on the other hand, does not remain constant along the separatrix. Specifically, it drops near the two X-points indicating that, close to these areas, hotter ions have exited the plasma. High energy ions with the right pitch angle will be in the ion loss cone\cite{chang2002x,ku2004property}. An interesting feature of the ion temperature is that the average temperatures below and above the outboard midplane are not the same: the region below the midplane has an average temperature which is $\SI{4.5}{\electronvolt}$ higher than the average temperature above. We can understand this difference in the context of the trapped ion banana orbits that we talked about in the particle fluxes section. The ions exiting the flux surface below the midplane are on the inward side of their banana orbit. They travel along a region of higher temperature therefore the temperature locally increases. Equivalently, when they re-enter above the midplane, they are coming from outside the LCFS where the temperature has dropped. Indicatively, a typical value for a banana orbit width near the edge is $\SI{6}{\milli\meter}$ and the temperature drop across this distance is about $\SI{2}{\electronvolt}$, accounting in part for the observed average temperature difference. Moreover, the re-entering ions could have originated from further away in the SOL than a typical banana orbit width, making the above estimate just a lower bound for the real temperature difference. Despite the above observations, the qualitative features of the equilibrium $E\times B$ ion heat flux are very similar to those of the equilibrium particle and electron heat $E\times B$ fluxes.    

The real difference between the two species heat flux behavior comes from the comparison between the relative sizes of equilibrium and turbulent heat fluxes. In the case of the electrons, turbulence dominates; neoclassical electron transport is negligible because of the small electron banana width. Thus, the local equilibrium $E\times B$ electron flux shown in Fig.~\ref{el_heat} will be further reduced by cancellation from bounce-orbit averaging.  Turbulence is the sole mechanism for net electron particle loss across the separatrix. It must compensate for the net loss of ions in order to maintain charge balance. Turbulent electron energy loss follows.     

In contrast, for the ions, it seems that the equilibrium process, set up by the neoclassical mechanism of trapped particle orbits, is indeed the dominant mode of ion loss, in this simulation, both for heat and particles. Both species LFS turbulent ballooning shapes are being interrupted in the high shear region. Although the ion turbulent heat flux never recovers and drops to zero, the electron turbulent heat flux picks up again after the initial dip. Moreover, in the region above the midplane, the electron turbulent heat flux is almost twice the size of the equivalent ion one, despite the big difference of the absolute temperature of the two species, with the ions being significantly hotter. This is a clear indication that, in the case of electrons, turbulence is absolutely necessary for the plasma to retain charge balance, whereas in the case of ions, the neoclassical processes are more effective. 

\begin{figure}
\includegraphics[width = \linewidth]{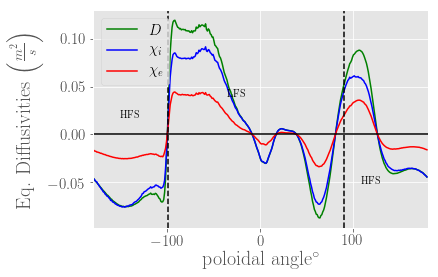}
\caption{Particle and heat, equilibrium $E\times B$ diffusivities.}
\label{eq_diff}
\end{figure}
\begin{figure}
\includegraphics[width = \linewidth]{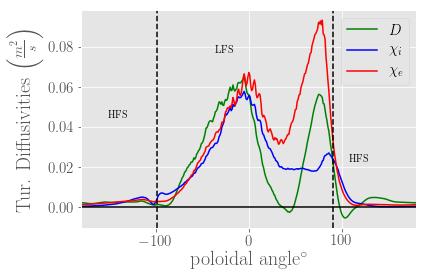}
\caption{Particle and heat, equilibrium $E\times B$ diffusivities.}
\label{turb_diff}
\end{figure}

In Figs.~\ref{eq_diff}-\ref{turb_diff} we present the effective diffusivities which are defined as $D = -\frac{\Gamma}{\partial n/ \partial x}$, $\chi_s = -\frac{Q_s}{\partial P_s/ \partial x}$ and we separate them again into equilibrium and turbulent contributions. The equilibrium diffusivities are presumably driven by neoclassical ion physics as discussed previously, but because of ambipolarity, equilibrium and turbulent diffusivities are fundamentally coupled. A first observation is that both equilibrium and turbulent diffusivities are of a similar order of magnitude, consistent with this coupling. 

In \rcite{2017APS..DPPTP11079}, methods developed for application to pedestal turbulence provide information about the modes present, based on the relative sizes of the transport channels. It is interesting to apply those concepts to our situation for qualitative insight. Here we find comparable turbulent diffusivities near the midplane in particle, electron heat and ion heat channels. This is suggestive of an underlying mode with an MHD character, possibly resistive ballooning or Kelvin Helmholtz in this electrostatic simulation. Near the top of the torus the electron diffusivity is dominant, which may suggest a mode driven primarily by the electron temperature gradient, or a mode with an electron drift character.  Indeed we find our turbulence has a frequency very close to the electron diamagnetic drift frequency. 

A final observation is that, based on the scale of the above diffusivities, the time scale for further profile evolution is much larger than the total time of the simulation. We have indeed verified that the profiles remain constant throughout, which is a clear indication of achievement of a quasi-steady turbulent state. 

\section{Quantitative Results}\label{sec:numerical}
\subsection{Separatrix-crossing Fluxes}
Here, we calculate the integrated currents and flux surface averages of the above fluxes and reach some conclusions regarding their relative importance in establishing the SOL width. 

First, we start with some definitions. The fluxes of the previous section are normal to the separatrix components of the total flux, i.e., $\Gamma_N = \frac{-B_R}{B_p}\left<n \cdot V^{E\times B}_{Z}\right> + \frac{B_Z}{B_p}\left<n \cdot V^{E\times B}_{R}\right>$. 
For a quantity $A$, the flux surface average of it is defined as: $\langle A \rangle = \frac{\int d\zeta d\theta \mathcal{J} A}{\int d\zeta d\theta \mathcal{J}}\,.$  

For the calculation of such integrals, we must know the Jacobian factor the inverse of which is given by the well known formula: 

\begin{align}
    \mathcal{J}^{-1} &= \nabla \Psi \cdot \nabla \theta \times \nabla \zeta \notag\\ 
    & = R B_p \hat{e}_{\Psi}\cdot \frac{1}{r \sqrt{1+\frac{1}{r^2}\frac{d^2r}{d\theta^2}}}\hat{e}_{\theta} \times \frac{1}{R} \hat{e}_{\zeta} \notag\\
    & = \frac{B_p}{r \sqrt{1+\frac{1}{r^2}\frac{d^2r}{d\theta^2}}}\,,
    \label{invJ}
\end{align}

with $r$ being the distance between $(R_o, Z_o)$ and the point on the flux surface. In the second step of the above calculation we have replaced the usual formula $\nabla \theta = \frac{1}{r} \hat{e}_{\theta}$, which holds in a circular geometry, with $\nabla \theta = \frac{1}{r \sqrt{1+\frac{1}{r^2}\frac{d^2r}{d\theta^2}}} \hat{e}_{\theta}$, which is an approximation for a general shape flux surface, provided the integrand is non-singular and single valued, that minimizes the integration error. Using this formula, the flux surface average (FSA) of a scalar $A$ is given by

\begin{equation}
    \langle A \rangle = \frac{\int d\zeta d\theta \frac{r \sqrt{1+\frac{1}{r^2}\frac{d^2r}{d\theta^2}}}{Bp} A}{\int d\zeta d\theta \frac{r \sqrt{1+\frac{1}{r^2}\frac{d^2r}{d\theta^2}}}{Bp}}\,.
\end{equation}

If we perform a flux-surface average on the continuity equation, we arrive at the expression: 
\begin{equation}
\frac{\partial \langle n \rangle}{\partial t} + \frac{1}{V^{\prime}}\frac{\partial}{\partial \Psi} V^{\prime}\langle \Gamma \cdot \nabla \Psi \rangle = 0\,,
\end{equation}
where $V^{\prime} = \int d\zeta d\theta \mathcal{J}$ and $\nabla \Psi = R B_p \hat{e}_{\Psi}\,$. Therefore, the physically relevant quantity is $\langle \Gamma \cdot \nabla\Psi \rangle$, which is calculated using the explicit formula:
\begin{equation}
\langle \Gamma\cdot \nabla \Psi \rangle = \frac{\int d\zeta d\theta r R \sqrt{1+\frac{1}{r^2}\frac{d^2r}{d\theta^2}} \Gamma}{\int d\zeta d\theta \frac{r \sqrt{1+\frac{1}{r^2}\frac{d^2r}{d\theta^2}}}{Bp}}\,.
\end{equation}
We can also define the surface integral of the flux across the LCFS. This should be interpreted as the particle or heat current, i.e., the number of particles per second that cross the separatrix due to the various fluxes. This will be denoted as $I_{\Gamma}$ and calculated using
\begin{equation}
    I_{\Gamma} = \int \Gamma R d\zeta \left(\sqrt{1+\frac{1}{r^2}\frac{d^2r}{d\theta^2}}\right)rd\theta\,,
\end{equation}
with similar expressions for the heat currents.

In Table~\ref{table:table_part}, we present the integrals over the whole separatrix. We have also kept for reference the numbers for flux from the magnetic drifts. In Tables~\ref{table:table_i_heat}-\ref{table:table_e_heat} we give the results for the integrated heat fluxes.

Here, we also define an averaging factor, $AF$, which is defined as $AF = \langle\Gamma \rangle/max\left(\Gamma \right)$ and shows the degree of ``cancellation" of each particular flux. If we assume that the density and potential are constant on a flux surface, then the FSA's of both the equilibrium $E\times B$ and the magnetic fluxes can be shown to be identically zero. Close to the edge of the plasma, as we saw in the previous section, the constancy of those quantities on the separatrix is violated hence, we don't expect the $AF$ to be exactly zero. Indeed, as we see in Table~\ref{table:table_cancellations}, the averaging factor for both the magnetic drift flux and the equilibrium $E\times B$ flux are very small but nonzero, indicating that the bulk of these fluxes are returning into the main plasma whereas, the $AF$ of the turbulent $E\times B$ flux is larger. $AF$'s of the heat fluxes are similar.
\begin{table}
\begin{tabular}{|c|c|c|c|}
\hline
\multicolumn{3}{|c|}{Particle Fluxes}\\ \hline
  & $\langle \Gamma \cdot \nabla \Psi \rangle$ & $ I_{\Gamma} $ \\ \hline
 Equilibrium  & 9.13 $\times 10^{18}$ & 1.17 $\times 10^{21}$ \\ \hline
 Turbulent  & 2.35 $\times 10^{19}$ & 3.02 $\times 10^{21}$ \\ \hline
 Magnetic  & 8.13 $\times 10^{18}$ & 1.04 $\times 10^{21}$ \\ \hline
\end{tabular}
\caption{Integrated Particle Fluxes.}
\label{table:table_part}
\end{table}

\begin{table}
\begin{tabular}{|c|c|c|c|}
\hline
\multicolumn{3}{|c|}{Ion Heat Fluxes}\\ \hline
  & $\langle Q_i\cdot \nabla \Psi \rangle$ & $ I_{Q_i} $ \\ \hline
 Equilibrium  & 3.17 $\times 10^{21}$ & 4.07 $\times 10^{23}$ \\ \hline
 Turbulent  & 1.05 $\times 10^{22}$ & 1.35 $\times 10^{24}$ \\ \hline
\end{tabular}
\caption{Integrated Ion Heat Fluxes.}
\label{table:table_i_heat}
\end{table}

\begin{table}
\begin{tabular}{|c|c|c|c|}
\hline
\multicolumn{3}{|c|}{Electron Heat Fluxes}\\ \hline
  & $\langle Q_e\cdot \nabla \Psi \rangle$ & $ I_{Q_e} $ \\ \hline
 Equilibrium  & 1.32 $\times 10^{21}$ & 1.69 $\times 10^{23}$ \\ \hline
 Turbulent  & 1.4 $\times 10^{22}$ & 1.80 $\times 10^{24}$ \\ \hline
\end{tabular}
\caption{Integrated Electron Heat Fluxes.}
\label{table:table_e_heat}
\end{table}

\begin{table}
\begin{tabular}{|c|c|c|}
\hline
\multicolumn{2}{|c|}{Averaging factor}\\ \hline
 & $\langle \Gamma\cdot \nabla \Psi \rangle$ \\ \hline
Equilibrium  &2.2\% \\ \hline
Turbulent  &8.6\%\\ \hline
Magnetic  &0.2\%\\ \hline
\end{tabular}
\caption{Averaging factors of Particle FSA Integrals.}
\label{table:table_cancellations}
\end{table}

A few comments about the relative sizes of those integrals are in order: In the particle fluxes and ion heat flux case, the size of the turbulent current is roughly three times larger than the equilibrium one. In the case of the electron heat fluxes though, the turbulent current is almost ten times larger than the equilibrium one indicating the importance of turbulence for the electrons, supporting the main argument of the previous section. 

\subsection{Particle Balance in the SOL}
In this sub-section, we explore particle balance in the SOL by comparing the flux of electrons exiting across the separatrix with the exhaust flux in the SOL, which is dominated by the parallel flows.  The competition between these cross-field and parallel transport processes is what sets the density SOL width.

Regarding the separate contribution of each flux to the setting of the SOL width the ratio of the currents doesn't tell the full story. Despite the fact that the turbulent flux is fully localized at the outboard midplane, contributing solely to the LFS SOL, the equilibrium fluxes have an important inward component at the HFS. Moreover, as we have mentioned in the previous sections, the particle $E\times B$ equilibrium flux cannot take electrons outside of the separatrix. Therefore, we start with the assumption that the current from the turbulent $E\times B$ particle flux is the total electron current exiting the separatrix. Now, we take a Gaussian box whose one side is the part of LCFS starting at the point where the $E \times B$ turbulent particle flux begins and ending where it ends. The top and bottom of this box are horizontal lines, starting at these two locations and extending outwards, deep inside the SOL. The extent of these lines is left to to be determined by the behavior of the normal fluxes on them.
The right hand side of the box is unspecified but we assume that no flux leaks from this side of the box to the wall. 

\begin{figure}
\includegraphics[width = 0.5\textwidth]{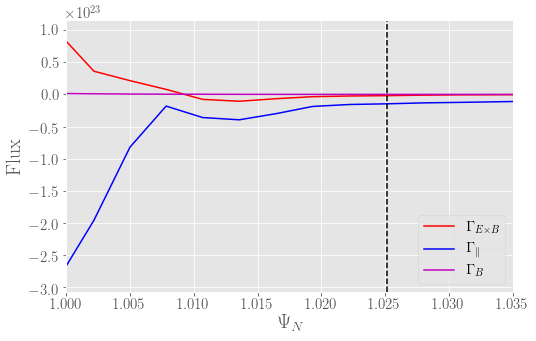}
\caption{Normal particle fluxes at the bottom cap of the Gaussian box. Dashed line indicates limit of integration. The horizontal axis is normalized poloidal flux.}
\label{bot_cap}
\end{figure}

\begin{figure}
\includegraphics[width = 0.5\textwidth]{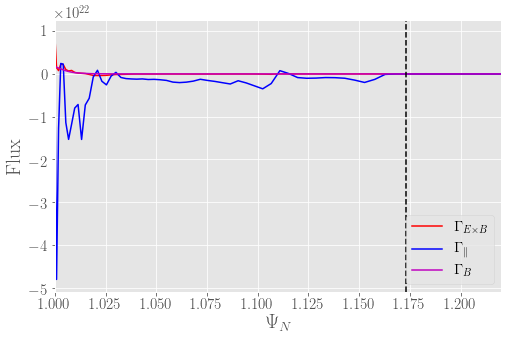}
\caption{Normal particle  fluxes at the top cap of the Gaussian box. Dashed line indicates limit of integration. The horizontal axis is normalized poloidal flux.}
\label{top_cap}
\end{figure}

In Figs.~\ref{bot_cap}-\ref{top_cap} we have plotted the normal component of the particle fluxes at the bottom and top caps of the Gaussian box respectively. The normal component of the particle flux includes contributions from the $E\times B$ flux, the magentic drift flux and, importantly, the parallel flows. We are not concerned with distinguishing the contributions from equilibrium and turbulent $E\times B$ therefore, we just present the total. An important note here is that, since we do not distinguish between equilibrium and turbulence, in the calculation of the $E\times B$ velocity we take $\frac{\partial \phi}{\partial \zeta} = 0$. We believe that the error introduced by such an approximation is minimal since, the $E\times B$ contribution to the normal fluxes is only important in a very narrow region close to the separatrix. There, the poloidal magnetic field is almost zero hence, $\frac{\partial \phi}{\partial \zeta} \approx B\cdot \nabla \phi = 0$. Generally, the parallel flux contribution is dominant everywhere and the only place where the other fluxes seem to be relevant is at a very narrow channel close to the LCFS. In both figures, the negative $Z$ direction is downwards, which means that for the top cap, negative fluxes are entering the box whereas for the bottom cap negative fluxes are leaving. 

In the case of the top cap, it seems that there is a well defined point where all fluxes drop to zero, around $\Psi_{N} \approx 1.16$. This seems to be a reasonable place to truncate the integration of the fluxes in order to find the total current crossing the top of the box. The total current thus found is $\SI{8.61e21}{s^{-1}} = \SI{1.38} {\kilo \ampere}$, entering the box.

For the bottom cap, things are not so clear cut. Although the fluxes coming from $E \times B$ and magnetic drifts do indeed drop to zero, the parallel flux seems to remain constant. This had to be expected since there is a significant degree of ionization happening in the SOL. It is very hard to precisely locate a point to truncate the integration, meaning to find where the SOL ends and what is left after that is simply ionization. We believe that a sensible point to stop our integration is at $\Psi_{N} \approx 1.025$ (shown in figure~\ref{bot_cap}). The total current found from this choice is $\SI{1.17e22}{s^{-1}} = \SI{1.87}{\kilo\ampere}$, exiting the box. If we extend the integration further out, this number will increase.

As we said above, the total electron current crossing into the box from the separatrix, has to be the one coming from the turbulent $E \times B$ flux given in Table~\ref{table:table_part} as $\SI{3.02e21}{s^{-1}} = \SI{0.48}{\kilo\ampere}$. If we compare this number to the total current found from the top and bottom of the boxes, we see that it accounts for $98 \%$. A direct conclusion that can be drawn from such a comparison is that indeed, as presumed in the Goldston model, the density channel is filled up almost entirely with turbulent electron flux. Of course, had we taken the limit of integration of the bottom cap further out, this percentage would have dropped however, it is reasonable to assume that this result would be severely contaminated by atomic processes that take place outside the LCFS and have nothing to do with turbulence and flows that cross the separatrix. 

Here, we need to note that even though we can safely claim that the electron contribution to the SOL channel is almost entirely turbulent, a similar statement can not be made for the ions. From the data that we have available, we have no way of isolating the turbulent and the neoclassical contributions since turbulence, in the ions' case, exists on top of a neoclassical background. 

If we repeat the previous calculation for electron heat fluxes, we find both the top and bottom caps of the box to have well defined limits where all fluxes go to zero. However, the comparison of the total electron heat current exiting the Gaussian box from the caps to the electron heat current entering from the separatrix, shows that the latter is more than twice the former. A plausible explanation for this result is that heat is dissipated in this volume due to various atomic processes, predominantly ionization. A detailed account of the relative strengths of such processes is beyond the scope of this paper. 



\section{Summary and Conclusions}\label{sec:conclusions}
In this paper we have analyzed the results of an XGC1 DIII-D simulation focusing on the fluxes exiting the separatrix. We reconstructed the $E\times B$ flux distinguishing between the equilibrium and the turbulent part. The poloidal patterns of the two particle fluxes were presented. For the equilibrium flux, we traced it's poloidal shape to the LCFS potential. We interpreted this potential structure as the result of ion drifts. Specifically, we could clearly distinguish the impact of the $\nabla B$-drifts and X-point losses on the potential's global maximum and minimum at the bottom X-point and top of the machine, respectively. For the poloidal potential variation at the LFS, we argued that it's due to the trapped ion orbits close to the edge that exit and re-enter the confinement. We constructed a simplified model for the angular distribution function of the banana orbits' inflection points and showed that the resulting ``trapped-particle" flux follows the shape of the equilibrium flux, to support our argument. Moreover, we noted that LFS edge electrons exhibit a significant departure from adiabaticity. We provided as reasons for this departure the fact that the turbulent frequency and the collision rate are comparable to the transit frequency. The turbulent flux pattern was also given. It appears to be strongly ballooning but with a significant region of suppression. We contended that this suppression is coming from the ion-neoclassically generated $E\times B$ shear which seems to be at the right level to affect the turbulence exactly at that region. 

We presented the $E\times B$ heat fluxes for ions and electrons at the separatrix. From the difference in size between the turbulent heat fluxes of the two species it is clear that turbulence is more important for the electrons since it is their only way of losing confinement. Ions are exiting the separatrix through neoclassical mechanisms, therefore, turbulent electron heat flux has to increase in order to maintain quasineutrality. The electrons were found to be isothermal whereas the ion temperature exhibits significant drops near the X-points, in accordance with the predictions of X-loss theory. The small temperature difference between the upper and lower halfs of the LFS can be, in part, explained by ion banana orbits exiting from a hotter and re-entering from a cooler part of the machine. The equilibrium and turbulent diffusivities of those fluxes were calculated, providing insight regarding the modes present. Numerical integrations were performed to calculate the flux-surface-average fluxes and the outgoing particle and heat currents from the separatrix. Lastly, we confirmed that anomalous electron diffusion accounts for almost all of the electrons filling the SOL particle channel.

Given the importance of the predictability of the SOL width for modern tokamaks, more state-of-the-art gyrokinetic simulations need to be analyzed to elucidate the complex physical processes that give rise to it. The interaction between the ion drifts and the turbulence seems to be critical for  setting the size of the channel and the study of the fluxes and flows at the edge provides illuminating insights into the nature of the system dynamics. In future publications we will examine simulations from different tokamaks and look into the characteristics of the turbulent ``blobs" from this and other simulations. These efforts will be directed towards obtaining a better understanding of the interaction of neoclassical and turbulent physics, and the extent to which the present results are applicable to other plasma and device regimes.

\newpage

\section*{Acknowledgements}
I. Keramidas Charidakos would like to acknowledge useful help with the data from Jugal Chowdhury at the early phase of this project.
We acknowledge computing resources on Titan at OLCF through the 2015 INCITE and the 2016 ALCC awards.
Work supported by the U.S. Department of Energy Office of Science, Office of Fusion Energy Sciences under Award Number DE-FG02-97ER54392 and by subcontract SO15882-C with PPPL under the U.S. Department of Energy HBPS SciDAC project.

\bibliographystyle{unsrt}
\bibliography{fluxbibliography}

\end{document}